\documentclass[journal]{IEEEtran}
\ifCLASSINFOpdf
   \usepackage[pdftex]{graphicx}
\else
\fi
\usepackage{amsmath}
\usepackage{amssymb}
\usepackage{algorithm}
\usepackage{algpseudocode}
\begin{document}
%
\title{Single-View Tomographic Reconstruction Using Learned Primal Dual}
%
%
%



\author{
    \IEEEauthorblockN{Sean Breckling,}
    \IEEEauthorblockA{Advanced Analytics, Nevada National Security Sites\thanks{This work was done by Mission Support and Test Services, LLC, under Contract No. DE-NA0003624 with the U.S. Department of Energy and the National Nuclear Security Administration’s Office of Defense Programs. DOE/NV/03624$\text{-}\text{-}$2280.}}
    
    \and
    
    \IEEEauthorblockN{Mathew Swan,}
    \IEEEauthorblockA{Advanced Analytics, Nevada National Security Sites}
    
    \and
    
    \IEEEauthorblockN{Keith D. Tan,}
    \IEEEauthorblockA{Computer Science Department, University of Nevada, Las Vegas}
    
    \and
    
    \IEEEauthorblockN{Derek Wingard,}
    \IEEEauthorblockA{Department of Physics, California State University, Long Beach}
    
    \and
    
    \IEEEauthorblockN{Brandon Baldanado,}
    \IEEEauthorblockA{Transformational Diagnostics \& Imaging, Nevada National Security Sites}
    
    \and
    
    \IEEEauthorblockN{Yoohwan Kim,}
    \IEEEauthorblockA{Computer Science Department, University of Nevada, Las Vegas}
    
    \and
    
    \IEEEauthorblockN{Ju-Yeon Jo,}
    \IEEEauthorblockA{Computer Science Department, University of Nevada, Las Vegas}
    
    \and
    
    \IEEEauthorblockN{Evan Scott,}
    \IEEEauthorblockA{Advanced Sources and Detectors, Nevada National Security Sites}
    
    \and
    
    \IEEEauthorblockN{Jordan Pillow,}
    \IEEEauthorblockA{Data Science, Nevada National Security Sites}
}

%
%

\markboth{Journal of \LaTeX\ Class Files,~Vol.~14, No.~8, August~2015}%
{Shell \MakeLowercase{\textit{et al.}}: Bare Demo of IEEEtran.cls for IEEE Journals}
%



\maketitle


\begin{abstract}
The Learned Primal Dual (LPD) method has shown promising results in various tomographic reconstruction modalities, particularly under challenging acquisition restrictions such as limited viewing angles or a limited number of views. We investigate the performance of LPD in a more extreme case: single-view tomographic reconstructions of axially-symmetric targets. This study considers two modalities: the first assumes low-divergence or parallel X-rays. The second models a cone-beam X-ray imaging testbed. For both modalities, training data is generated using closed-form integral transforms, or physics-based ray-tracing software, then corrupted with blur and noise. Our results are then compared against common numerical inversion methodologies.

\end{abstract}


%
\IEEEpeerreviewmaketitle

\section{Introduction}
\IEEEPARstart{H}{igh-Speed} flash radiographic imaging is a mainstay diagnostic in materials science and shock physics. Given the amount of light required to sufficiently illuminate a fast-moving hydrodynamic scene grows proportionally with that speed \cite{Introduction:versluis2013}, a tomographic imaging test bed capable of capturing a hydrodynamic target scene with several sufficiently well-resolved viewing angles would require each axis to be triggered at the same instant. As a result, shock physicists have traditionally designed experiments to proceed in an axially-symmetric trajectory. This configuration not only simplifies the mathematical model of the experiment, but it also reduces the number of projections sufficient for a complete tomographic reconstruction to as few as one \cite{dooley1961calculation, houwing2005abel, benuzzi2006laser}.

\begin{figure}
    \centering
    \includegraphics[width = 3.5in]{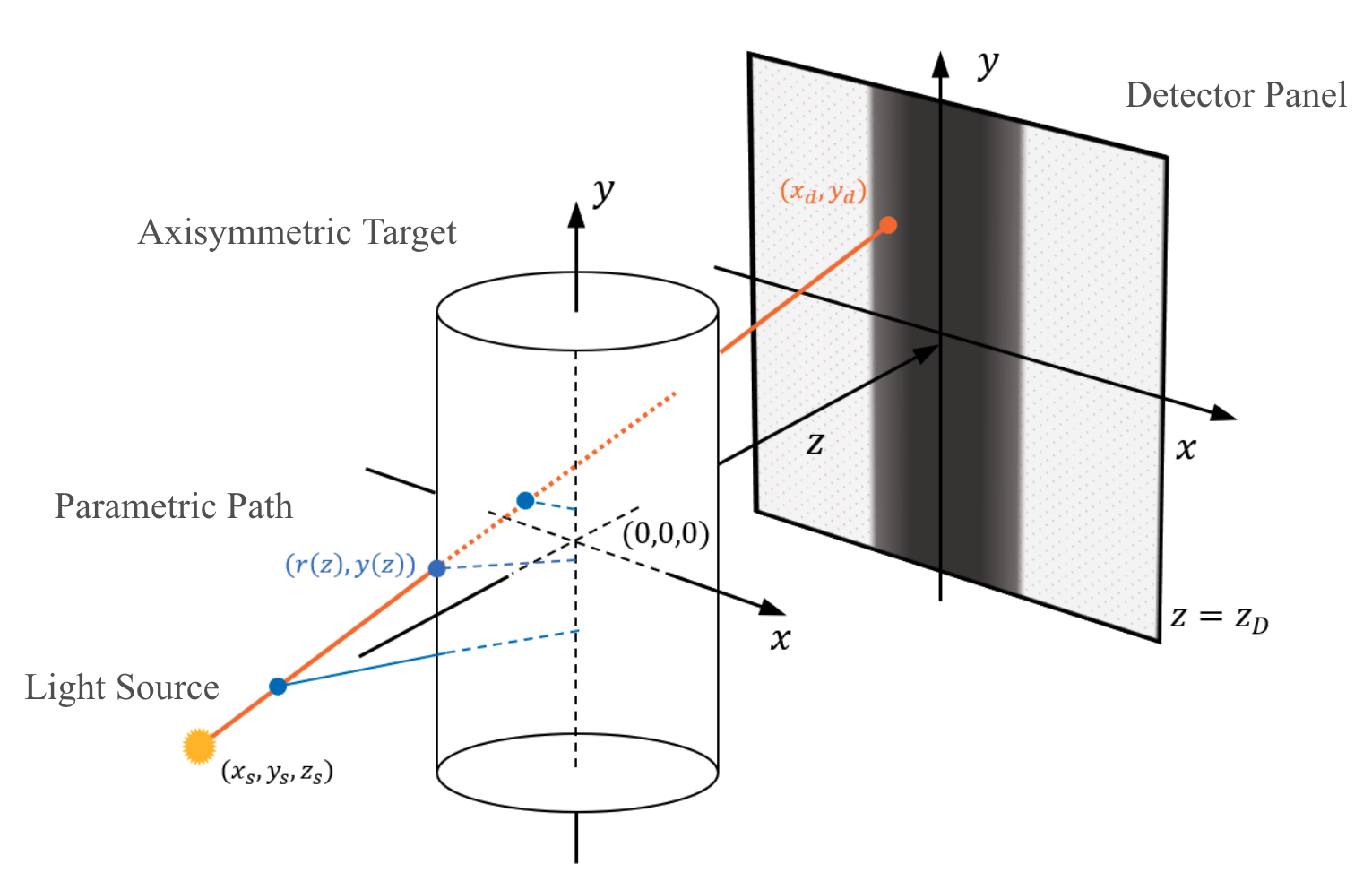}
    \caption{Depicted is a simplified model of an arbitrary single-view X-ray imaging apparatus. The light source is located at $(x_s, y_s, z_s)$ in cone-beam applications and $(x_s, y_s, -\infty)$ under the parallel-beam assumption. The origin is located on the axis of symmetry, and the detector panel is assumed to be on the $z = z_D$ plane. The orange line denotes a particular photon path from the source to detector, and the blue lines depict various positions along that path, as parameterized in cylindrical coordinates.}
    \label{fig:system}
\end{figure}

Let the volumetric density of a 3-dimensional imaging target  $\hat{u}(x,y,z)$ be axisymmetric along the $y$ axis, i.e. there exists some $u(\sqrt{x^2 + z^2},y) = \hat{u}(x,y,z).$ Let us further assert that the shortest path between the light source centered at the point $(x_s, y_s, z_s)$, and detector plane $z=z_d$ is parallel to the $z$ axis. Given an axisymmetric reconstruction space $X$, and data space $Y,$ the integral transform
\begin{equation}\label{eqn::cone}
d(x_i,y_j) = \mathcal{A} u(r,y) = \int_{z_s}^{z_d} u(r(z), y(z)) dz,
\end{equation}
denotes a forward projection model $\mathcal{A}:X \rightarrow Y$ from $u \in X$ to $d \in Y.$ A diagram can be seen in Figure \ref{fig:system}.

It is clear from inspection that the axial symmetry assumption on $X$ imposes a sampling singularity along the $r=0$ axis. It is for this reason that, outside of ideal cases, solving inverse problems associated with Equation \ref{eqn::cone} with sufficient regularity remains a considerable challenge. Direct numerical optimization schemes employed on the usual least squares loss model
\begin{equation} \label{eqn:lsq}
    \min_{u \in X} ||\mathcal{A}u - d||_{L^2(Y)}^2
\end{equation}
are often fraught with instabilities stemming from errors in approximating  $\mathcal{A}$. In practice, among the most-common culprits include poor model parameterization, or stochastic noise sources; particularly near the axis of symmetry \cite{Dasch:92}. These difficulties are compounded when the X-ray light source is diverging; i.e. the \emph{cone beam} modality. Geometrically-accurate axisymmetric cone-beam models are often poorly conditioned, and can require a significant amount of work to stabilize, even with modern methods.

These difficulties have motivated the study and implementation of regularized loss models of the form
\begin{equation}\label{eqn:reg}
  \min_{u \in X} \left[ \frac{\lambda}{2}||\mathcal{A}u - d||_{L^2(Y)}^2 + \mathcal{R}(u) \right], \ \lambda \geq 0
\end{equation}
where $\mathcal{R}$ is the regularization penalty functional. Examples include Tikhonov \cite{daun2005solution}, Total Variation (TVmin) \cite{asaki2006abel}, higher-order TVmin \cite{chan2015high}, and $L1/L2$ of the gradient \cite{breckling2023box}. Such techniques are typically approached by solving their respective loss models using a direct numerical optimization technique, e.g. the Primal Dual approach of Chambolle \& Pock \cite{chambolle2011first}. This approach has been studied extensively in literature, even for single-view tomographic reconstruction \cite{zhang2018stability}.

The use of Convolutional Neural Networks (CNNs) has shown promise in side-stepping these and similar difficulties, typically through judicious selection of relevant training data \cite{kim2019extreme}. Existing approaches typically fall into two distinct categories. Direct-to-image architectures, e.g. \cite{zhu2018image} or \cite{germer2023limited}, are common in literature and endeavor to learn the entire inverse mapping from scratch, often necessitating wide and deep fully-connected networks. While powerful, such "black box" methods are expensive to train, and prone to over-fitting specific acquisition geometries. Conversely, post-processing schemes like Filtered Back Projection Convolutional Neural Network (FBPConvNet) \cite{jin2017deep} or High-frequency Enhanced and Attention-guided Learning networks (HEAL) \cite{li2024heal} act as image-domain denoisers, which are trained to repair artifacts and errors common to fast / direct spectral solvers like Feldkamp-Davis-Kress (FDK) \cite{feldkamp1984practical}. However, by decoupling the restoration process from the physics of the acquisition model, these methods lack a strong mechanism to verify data consistency, which increases the risk of hallucination.

Herein we consider a Learned Primal Dual (LPD) architecture for single-view tomographic reconstruction problems. This method has recently shown promise in limited-angle and extreme few-view modalities for tomographic reconstruction, particularly compared to direct numerical optimization-based methods in the severely under-determined cases \cite{Adler2019}. The LPD scheme is a so-called \emph{unrolled} scheme, as it is constructed by substituting the eponymous Primal and Dual steps with shallow CNNs. These shallow networks are coupled in sequence, with the forward model $\mathcal{A}$ and adjoint $\mathcal{A}^*$ applied between the primal and dual steps, iteratively enforcing the physical acquisition model constraints. This architecture significantly reduces the network complexity compared to monolithic methods, and ensures reconstructions remain consistent with the projection data, minimizing the hallucination risks common to the post-processing approaches. These are key safeguards for axisymmetric tomography, given the severity of the acquisition model assumptions.

The use of NNs as part, or all, of the inversion procedure in single-view modalities is not new. Multilayer Perceptron (MLP) models were implemented with success as early as 1992 \cite{bishop1992neural}, with NN-based methods arriving in the literature by 2002 \cite{Ma2002}. However, the rapid growth of available software tools, computational resources, and broad interest in machine learning (ML) has resulted in a significant uptick in new studies  \cite{RODRIGUEZ2021119011, Duann2023, liu2023physics}. 

We consider two use-cases of the axisymmetric tomography problem. We begin with the usual parallel beam assumption common to inverse problems in astronomy \cite{zhang2023reconstructing}, pyrometry \cite{dreyer2019improved}, and X-ray radiography at low-divergence / compressed scientific light sources \cite{theocharous2019use}. Under this parallel beam assumption, one can reduce the general forward operator (\ref{eqn::cone}) to the Abel transform
\begin{equation}\label{eqn::Abel}
    d(x_i,y_j) = \mathcal{A} u(r,y_j) = 2\int_{|x_i|}^\infty \frac{u(r,y_j) dr}{\sqrt{r^2 - x_i^2}}.
\end{equation}
In contrast, our second modality presents strong geometric magnification, and cone-beam effects common to flash X-ray systems like Cygnus testbed within the Principal Underground Laboratory for Subcritical Experimentation (PULSE) at the Nevada National Security Sites (NNSS) \cite{smith2005cygnus, smith2007}, Dual-Axis Radiographic Hydrodynamic Test (DARHT) at Los Alamos National Laboratory \cite{burns2002overview} , or the Z Pulsed Power Facility (Z Machine) at Sandia National Laboratories \cite{webb2023radiation}. 

The remainder of this paper is outlined as follows. We introduce and discuss the LPD scheme in Section \ref{sec::lpd}, detailing network architecture, as well as the training and evaluation metrics. In Section \ref{sec::demo} we demonstrate the efficacy of LPD against competing techniques in both the parallel and cone beam modalities. Lastly, we provide concluding remarks in Section \ref{sec::conc}.


\section{Implementing the Learned Primal Dual Method}\label{sec::lpd}
The nonlinear extension to the Chambolle-Pock scheme is given in Algorithm \ref{alg:cap}. The key Primal and Dual steps are given in terms of proximal operators of the form
$$prox_{\tau\mathcal{G}}(u) = argmin_{v \in X}\left(\mathcal{G}(v) + \frac{1}{2\tau} ||u-v||\right),$$
where the real-valued step parameter $\tau > 0$, and $\mathcal{G} : X \rightarrow Y.$ Letting $\mathcal{G}(\cdot)$ denote the negative data log-likelihood $\mathcal{L}(\cdot, d)$ gives rise to the following algorithm.

\begin{algorithm}
\caption{Nonlinear Primal Dual Hybrid-Gradient}\label{alg:cap}
\begin{algorithmic}
\Require$\sigma, \tau > 0 $ s.t. $ \ \sigma \tau ||\mathcal{A}||^2 < 1, \ \gamma \in [0,1]$ with  $u_0~\in~X,$ and $d,h_0~\in~Y.$

\For{$i = 1, \ldots, n_{stop}$} 
\State $h_{i+1} \leftarrow prox_{\sigma \mathcal{G}^*}\left( h_{i} + \sigma \mathcal{A}(\overline{u}_i)\right)$ 
\State $u_{i+1} \leftarrow prox_{\tau \mathcal{R}}\left( u_{i} - \tau \mathcal{A}^* (h_{i+1})\right)$ 
\State $\overline{u}_{i+1} \leftarrow u_{i+1} + \gamma (u_{i} - u_{i+1})$
\EndFor
\end{algorithmic}
\end{algorithm}

\begin{figure*}[t]
    \centering
    \includegraphics[width=6in]{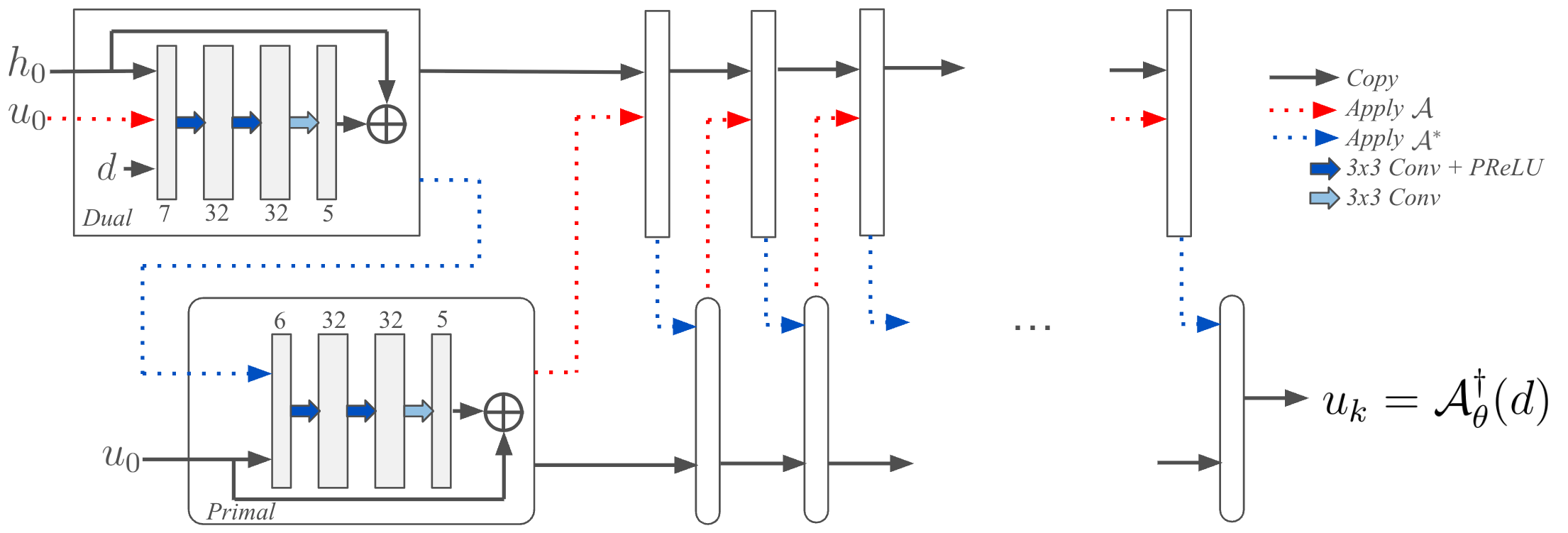}
    \caption{A flow diagram illustrating the LPD network architecture that highlights the structure of a single iteration. The network consists of two CNN blocks that respectively approximate the dual and primal steps of Algorithm \ref{alg:lpd}. The forward model ($\mathcal{A}$) and its adjoint ($\mathcal{A}^*$) are used to enforce data consistency and pass information between the dual and primal domains. The numbers shown (e.g., 7, 32, 5) represent the number of feature channels in the convolutional layers of each network block.  \ref{alg:lpd}.}
    \label{fig:LPD_diagram}
\end{figure*}

The LPD scheme is often referred to as an unrolled scheme. Notice that Algorithm \ref{alg:cap} iterates the $prox_{\sigma \mathcal{G}^*}$ and $prox_{\tau \mathcal{R}}$ procedures $n_{stop}$ times with sequentially updated data. With LPD, we instead enumerate and train $n_{stop}$ independent CNNs $\Gamma_{\theta_i^q}$ and $\Lambda_{\theta_i^p}$ respectively, where $p,q \in \mathbb{N}$ denote the lengths of hyperparameter vectors. A functional diagram of Algorithm 2 is included in Figure \ref{fig:LPD_diagram}.

\begin{algorithm}
\caption{Learned Primal Dual}\label{alg:lpd}
\begin{algorithmic}
\Require$u_0 \in X,$ and $d, h_0 \in Y$

\For{$i= 1, \ldots, n_{stop}$} 
\State{$h_i \leftarrow \Gamma_{\theta_i^d}(h_{i-1}, \mathcal{A}(u_{i-1}), d)$}
\State{$u_i \leftarrow \Lambda_{\theta_i^p}(u_{i-1}, \mathcal{A}^*(h_i))$}
\EndFor
\end{algorithmic}
\end{algorithm}

\subsection{Training Data}
It is remarkably uncommon for flash X-ray testbeds to have a catalog of radiographic image data of sufficient size to train a modest CNN. These data are often stored as small collections of images from disparate and highly specific experimental campaigns, each with unique prevailing settings and parameters. As such, since the creation of a unified and consistent dataset necessary for robust CNN training is not practical, we must rely on physically-relevant surrogate models.   

Given a particular surrogate imaging apparatus model $\mathcal{F} : X \rightarrow Y$, each scene in $X$ is constructed by combining one or more randomly-selected axisymmetric target objects, and assigning a volumetric density distribution. The corresponding projection image in $Y$ is then generated by calling the forward model $\mathcal{F},$ along with any associated distortion model (e.g. blur, and noise) $K$. Forward models denoted by $\mathcal{F}$ are either an analytic, closed-form model akin to those denoted below in Table \ref{tab:test_functions}, or an absorption image generated using a physics ray-tracing engine like gVirtualXray (gVXR) \cite{vidal2024x, vidal2025x}, Monte Carlo N-Particle (MCNP) \cite{mcnp}, or HADES \cite{aufderheide2010inclusion}. We note that models $\mathcal{F}$ are only used to generate training data, and are distinct from the discretized models discussed in the following subsection. 

\subsection{Discretized Forward Models}\label{sec:DFM}
The usual approach to discretizing single-view, axisymmetric projection operators is not dissimilar from those taken in the more-general Radon transform case \cite{averbuch2008framework}, apart from the reduction from a 3D target space to a 2-dimensional cylindrical space reflected in (\ref{eqn::cone}). We consider the complete radiographic projection in both modalities. This choice is a departure from typical discretizations of (\ref{eqn::cone}), given that the idealized reconstruction problem is fully determined with precisely half of a complete projection of the cylindrical scene. In practice, the additional samples in the over-determined problem can help suppress the effects of measurement noise in the reconstruction process, but are often handled post-hoc. 

\begin{figure}
    \centering
    \includegraphics[width = 3.4in]{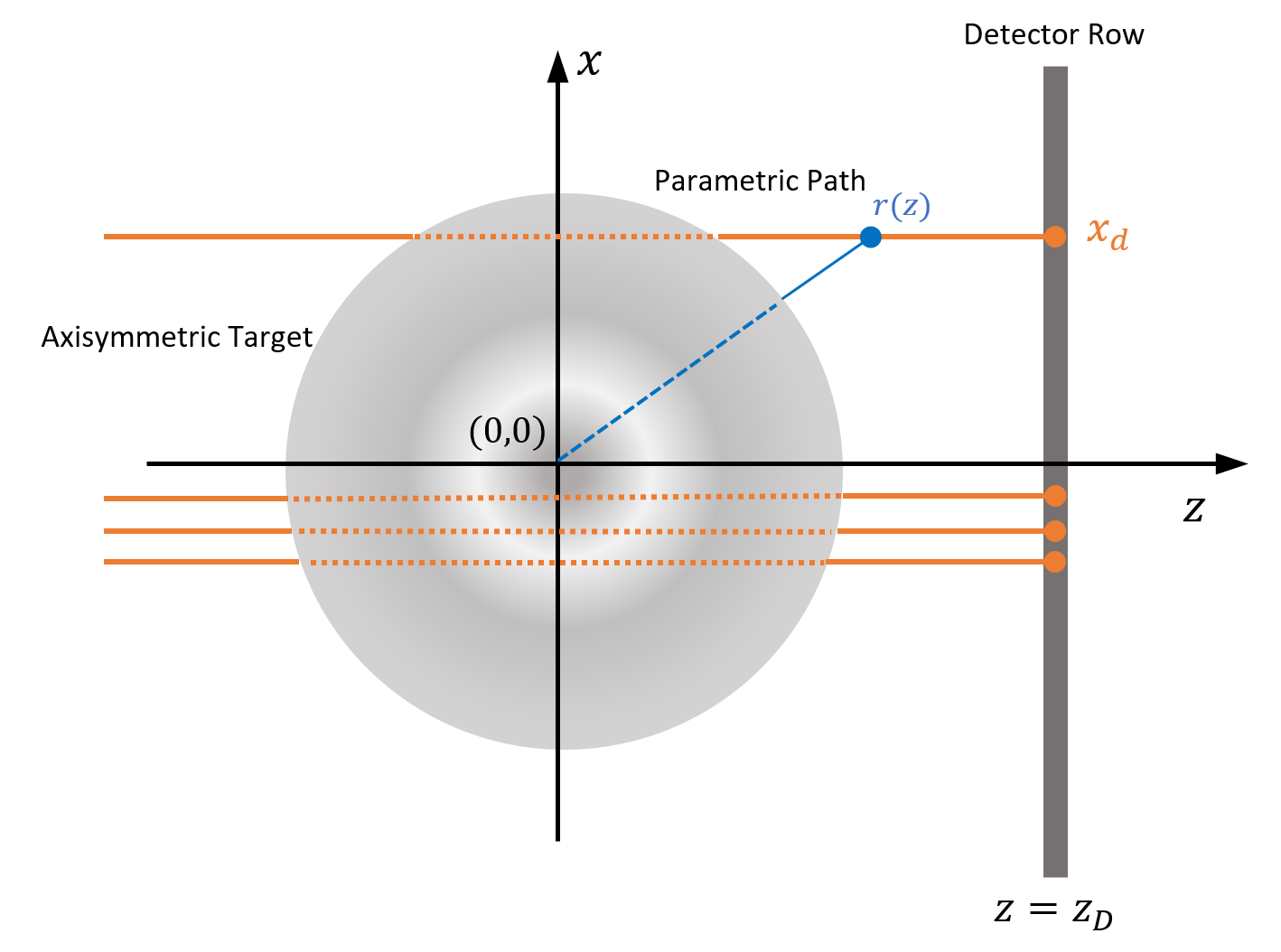}
    \caption{This diagram depicts a 2D slice of a cylindrical target imaged under the parallel beam modality. If it as assumed that the axis of symmetry is parallel with the $y-$axis of the detector panel, each row of pixels on the detector can be treated independently.}
    \label{fig:abel2Dgeo}
\end{figure}

For the parallel-beam modality, we discretize $\mathcal{A}$ using the so-called Onion-Peeling method \cite{hanson1993special} 
\cite{Dasch:92}. Given a maximal radius for the reconstruction cylinder $r_{max} > 0$, and a fixed number pixels $m \in \mathbb{N},$ the standard discretization of \eqref{eqn::Abel} results in a square matrix  $B \in \mathbb{R}^{m \times m}$ where 
$$r_{max} = (m+ (1/2))\Delta r.$$ 
Given that we intend to consider the entire ROI, we discretize the $x-$axis similarly, such that $x_{max} = -x_{min} = r_{max}.$ Since each row of pixels in the ROI is handled independently, as depicted in Figure \ref{fig:abel2Dgeo}, the resulting linear system mapping along any row $0 \leq j \leq m$ row $X_j \rightarrow Y_j$ is given by $A = [B, B]^\intercal$. To extend this discretization to the entire ROI, we write
$$\mathcal{A} = A \otimes I_{n},$$
where $n > 0$ is the number of rows present in the ROI, and $\otimes$ is the Kronecker product.

In the cone-beam modality we proceed by enumerating each possible ray path $L_{\bf i}$ from the fixed X-ray source to a particular pixel in the $n_{row} \times n_{col}$-pixel ROI as $0~<=~{\bf i}~<~n_{row}~\times~n_{col}.$ Similarly, we enumerate each annulus $\Gamma_{\bf j}$ within the reconstruction cylinder as $0 \leq {\bf j} < m_{row} \times m_{col}$. Put simply, the first-order, sparse forward-operator is given by the lengths of each intersection
$$\mathcal{A} = \left\lbrace a_{{\bf i}, {\bf j}} \right\rbrace = |L_{\bf i} \cap \Gamma_{\bf j}|.$$ 

\subsection{Network Architecture}
All models were implemented and trained using the PyTorch deep learning framework \cite{paszke2019pytorch}. Each convolutional layer within the Primal and Dual CNN blocks, as depicted in Figure \ref{fig:LPD_diagram}, utilized a 3x3 kernel followed by a Parametric Rectified Linear Unit (PReLU) activation function \cite{he2015delving}. The final layer of each block uses a linear activation. PREeLU is often selected over standard ReLU to allow the network to learn and adapt the negative slope; as was the case in \cite{Adler2019}. In the usual way, these networks are trained end-to-end using Adam \cite{adam2014method} against an MSE loss, and a fixed learning rate of $10^{-4}.$

All training is performed using one node of Lawrence Livermore National Laboratory's CZ-Matrix GPU cluster. Each node consists of four Nvidia H100 GPUs, with a total of 320GB of VRAM.

\subsection{Evaluation Metrics}\label{sec::eval_metrics} 
In the sections that follow, we utilize several common benchmarking metrics-- namely, the Mean-Square Error (MSE), Normalized Mean-Square Error (NMSE), the Structural Similarity Index Metric (SSIM) \cite{wang2004image}, and Pseudo Signal to Noise Ratio (PSNR). Given $N$, the total number of pixels in the image space $X$, a known non-zero ground truth $u \in X$ and approximation $v \in X$, we write
$$\text{MSE}(u,v) := \frac{||u-v||_X^2}{N},$$
$$\text{NMSE}(u,v) := \frac{||u-v||_X^2}{||u||_X^2},$$
$$\text{PSNR}(u,v):= 10\log_{10}{\left(\frac{||u||_\infty^2}{MSE(u,v)}\right).}$$
Given two small positive constants $c_1,$ and $c_2,$ as well as the means $\mu_u, \mu_v$ standard deviations $\sigma_u,\sigma_v$ and covariance $\sigma_{uv}$,
$$SSIM(u,v) := \left(\frac{\mu_u \mu_v + c_1}{\mu_u^2 + \mu_v^2 + c_1} \right)\left(\frac{2\sigma_{uv} + c_2}{\sigma_u^2 + \sigma_v^2 + c_2}\right).$$
Typically, the final SSIM value is computed by averaging the SSIM of sliding windows of sub-regions of the complete images $u,v \in X.$ 

\section{Numerical Demonstrations}\label{sec::demo}
In this section we discuss our implementations of Algorithm \ref{alg:lpd} through two complementary studies. Our first considers the simplest, fully analytic parallel beam modality defined below in Table \ref{tab:test_functions}, augmented with a modest amount of noise and blur. The second study considers a practical scenario where we train a LPD-based inversion tool for use at a particular X-ray imaging testbed. 

\subsection{Parallel-Beam Modality}
Let $M$ and $L$ denote arbitrary unit mass and length scales. All synthetic data for this modality is constructed by first generating a random collection of axisymmetric target scenes $u$ using combinations of volumetric density distributions. Idealized forward projections $\mathcal{F}(u)$ are then generated directly, using the corresponding transform functions in Table \ref{tab:test_functions}. We further restrict our training data corpus to the distributions in rows 1, 2, and 5 for Table \ref{tab:test_functions}.

{\renewcommand{\arraystretch}{2}%
\begin{table}[H]
    \centering
        \caption{Five axisymmetric volumetric density distributions and their analytic projections. Characteristic functions over real-valued intervals $[\delta_1,\delta_2] \subset \mathbb{R} $ are given by $\Pi_{[\delta_1,\delta_2]}$. }
    \label{tab:test_functions}
    \begin{tabular}{|c|c|c|}
       \hline 
        $Eqn.$ & $u(r)$             & $\mathcal{F}(u)(x)$ \\
         \hline                               
        1 & $ \Pi_{[0,\delta]}(r) $   & $2\sqrt{\delta^2 - x^2}\Pi_{[-\delta,\delta]}(x) $ \\
        2 & $ \sqrt{\delta^2 - r^2}\Pi_{[0,\delta]}(r) $ & $\frac{\pi}{2}(\delta^2 - x^2)\Pi_{[-\delta,\delta]}(x) $ \\
        3 & $ (\delta^2 - r^2)\Pi_{[0,\delta]}(r) $ & $\frac{4\pi}{3}(\delta^2 - x^2)^{3/2}\Pi_{[-\delta,\delta]}(x) $ \\        
        4 & $ (\delta^2 - r^2)^{3/2}\Pi_{[0,\delta]}(r)$& $\frac{3\pi}{8}(\delta^2 - x^2)^2\Pi_{[-\delta,\delta]}(x) $\\
        5 & $e^{-r^2 / \sigma^2}$ & $ |\sigma| \sqrt{\pi} e^{-x^2/\sigma^2}$\\
        \hline 
         
    \end{tabular}
\end{table}
}

Though our intended problem assumes an imaging detector of size $500 \times 256$ pixels $d_{i,j}$ with $\Delta x=\Delta y = 1/128$, our training data will be limited to a restricted detector region of size $20 \times 256$ pixels. This is a departure from the usual approach for parallel-beam modalities, where it is more common to simply treat each row of pixels independently. However, given that imaging testbeds often endure localized pixel-to-pixel correlations from blurring or other optical effects that extend beyond individual rows, we are deliberately constructing this LPD-based solver to consider data in minimal multi-row chunks sufficiently large to capture those distortions.

Blurring is accomplished using the usual Gaussian point-spread function (PSF)
$$K(r) = \exp\left(-\frac{\ln(2)}{4\Delta x^2}r^2\right),$$ which has a full-width at half-max (FWHM) of $2\Delta x$. This blur applied to each projection $F(u)$ using the fast Fourier-transform (FFT). Given a nominal intensity $I_0 = 10^5$, noise is added pixel-wise according to 
$$\hat{d}_k \sim P(I_0  e^{-K\star \mathcal{F}(u_k)} ), \ \text{then} \ \ \  d_k  = -\ln(\hat{d}_k / I_0)$$
where $\star$ denotes convolution, and $P$ a Poisson generator. 

\subsubsection{Training Details} 
\begin{figure}[H]
    \centering
    \includegraphics[width=3.4in]{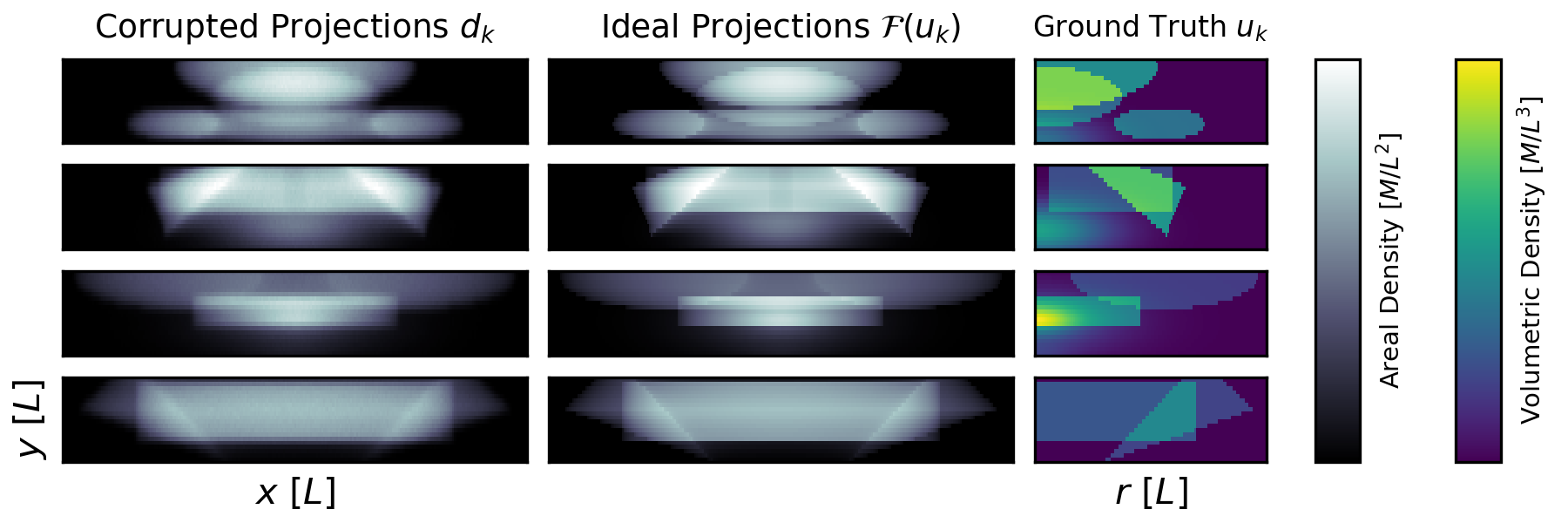}
\caption{Four example training pairs $(d_k,u_k)$ depicted on shared (qualitative) areal and volumetric density axes, along with a noise and distortion-free projection $\mathcal{F}(u_k)$ for comparison.}
    \label{fig:four_samples}
\end{figure}

A total of 6250 distinct target scenes $u$ and projections $\mathcal{F}(u)$ were generated. We established our training set with the first 5,000 pairs, with the evaluation set consisting of the remaining 1,250. This results in the usual 80/20 split, prior to adding blur and noise. Further, we implement a $5\times$ data augmentation to stabilize training and prevent over-fitting \cite{shorten2019survey}. Though we apply the noise and blur distortions for each of the 6,250 ideal pairs $(u,\mathcal{F}(u))$, we repeat this process four additional times for each of the 5,000 pairs in the training set. The result in an augmented training set of 25,000 pairs $(u_k,d_k).$ Example training pairs and clean projections are shown in Figure \ref{fig:four_samples}. The model was trained for 20,000 epochs with a batch size of 512 pairs $(u_k,d_k)$, resulting in a final MSE loss of 1.62E-2.

\subsubsection{Results}
Though we trained the LPD scheme on images with a 20-pixel row space, as mentioned above, we focus both our qualitative and quantitative comparisons on data with a much larger 500-pixel row space. In order to test the trained model's ability to generalize beyond its direct training domain, this set of 1,250 larger images contain a larger quantity of synthetic obstructions, and are assigned volumetric density distributions from all five rows of Table \ref{tab:test_functions}. 

\begin{figure}[H]
    \centering
    \includegraphics[width = 3.4in]{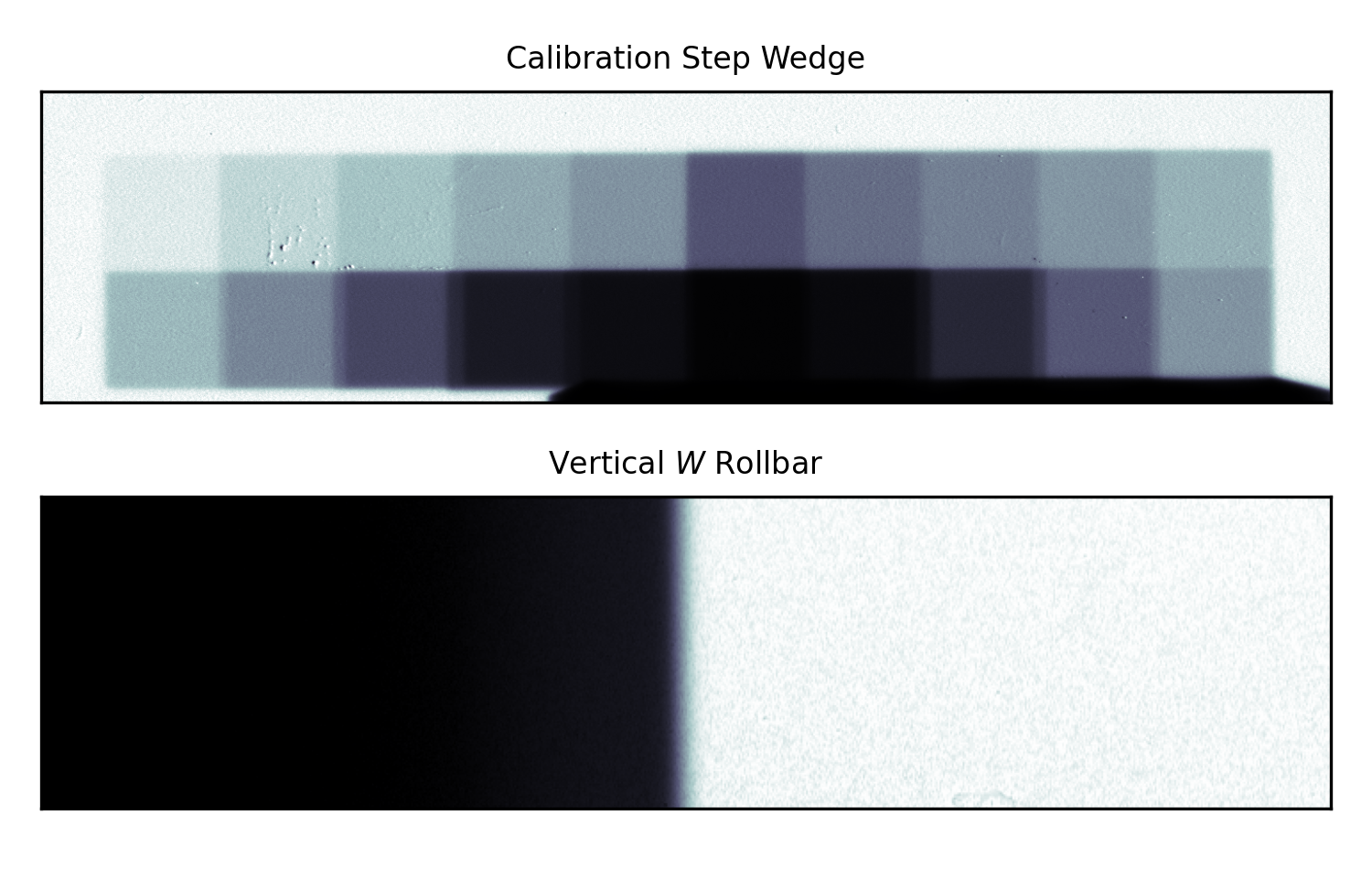}
    \caption{The top figure depicts a transmission radiograph cropped around a step wedge calibration device. Some small defects in the scintillating detector panel can be observed. These regions were avoided for the noise model study. The bottom depicts a cropped transmission radiograph of a vertical Tungsten rollbar.}
    \label{fig:calibration}
\end{figure}

\begin{figure*}[t]
    \centering
    \includegraphics[width=6in]{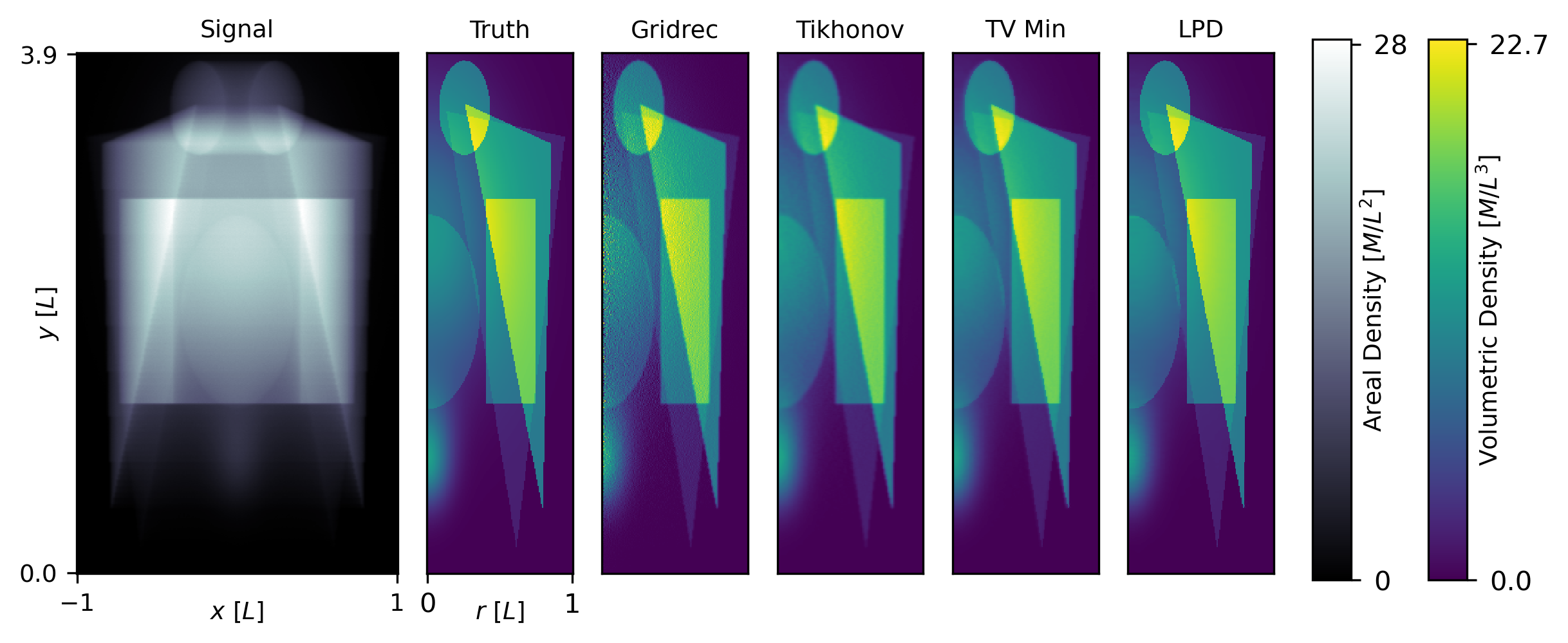}
\caption{From left to right, we start with a full-size, blur and noise-distorted image $d_k$, and the corresponding ground truth $u_k.$ The four results that follow are a selection of numerical inversions of $d_k$, varying only the methodology. The third image is computed using the Gridrec method, followed by Tikhonov and TV-min. The last image is the result of the row-wise LPD scheme.}
    \label{fig:compare_with_direct}
\end{figure*}

\begin{figure}
    \centering
    \includegraphics[width=3.4in]{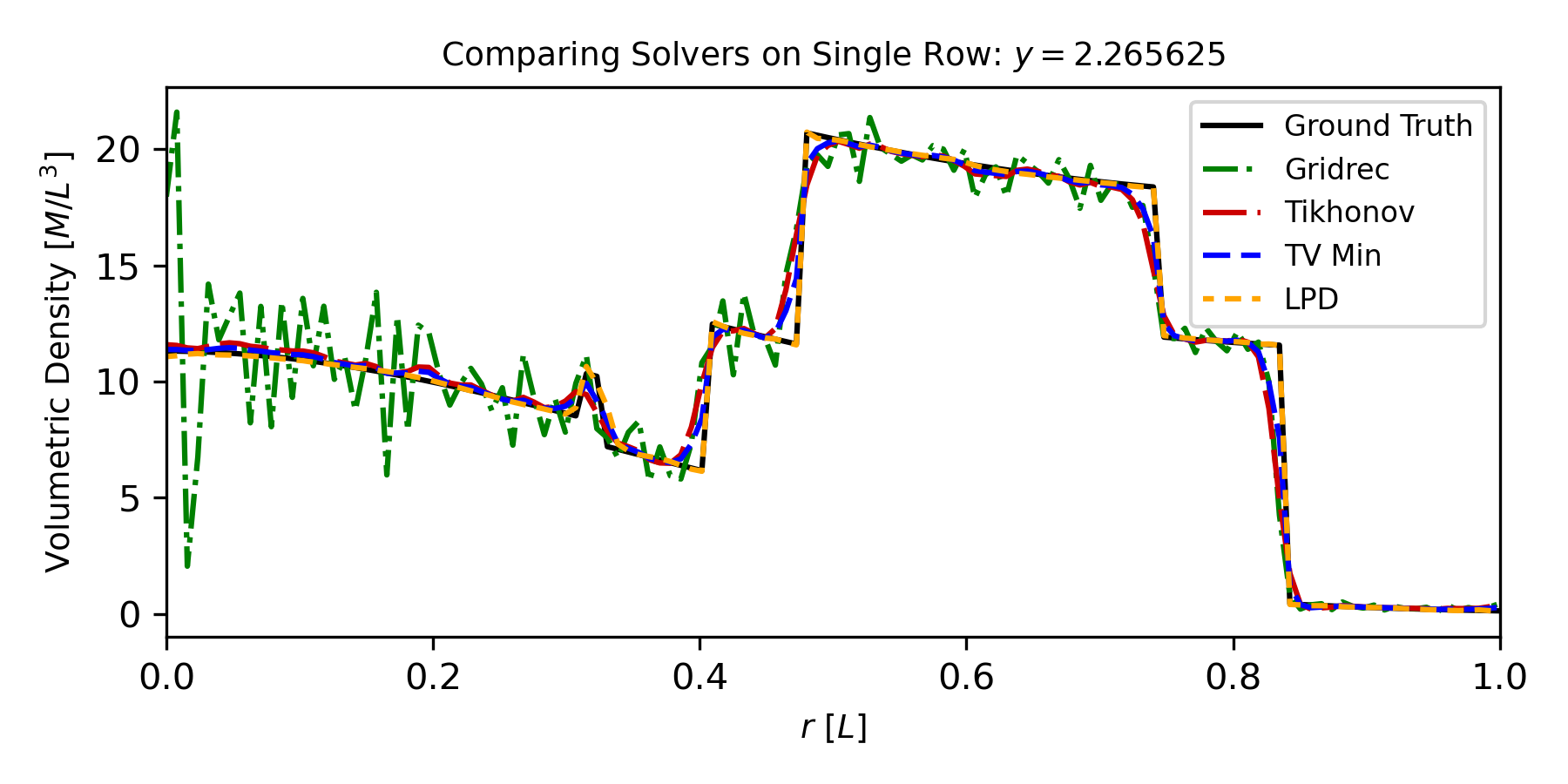}
\caption{Comparing the four reconstruction methods over a single line-out, overlaid upon the ground truth.}
    \label{fig:one_line_parallel}
\end{figure}

In order for our LPD scheme to accommodate images with the larger row-spaces, we implement a simple sliding window approach replacing only the tenth row of the 20-pixel window. In both Figures \ref{fig:compare_with_direct} and \ref{fig:one_line_parallel} we compare the LPD results to three common inversion techniques. First, we use a fast, unregularized baseline Gridrec \cite{dowd1999developments, gursoy2014tomopy}. Second, we use a direct minimization of \eqref{eqn:reg} using Tikhonov Regularization. Third, the TV-min using Algorithm \ref{alg:cap}. The regularization parameters $\lambda$ for both the Tikhonov and TV-min methods were identified using a grid-search method over the training set.

The average NMSE, SSIM, and PSNR were computed over each of the 1,250 images; the results were compiled in Table \ref{tab:parallel_results}. At first glance, these results, as well as Figures \ref{fig:compare_with_direct} and \ref{fig:one_line_parallel} show Algorithm \ref{alg:lpd} strongly out-performing the selected inversion schemes. The reason for such a comparatively strong result is our inclusion of numerical blurring and noise distortions to the training data. Neither blur nor Poisson noise are accounted for in the discrete forward model discussed in Section \ref{sec:DFM}, therefore direct minimization of the least squares error \eqref{eqn:lsq} could not meaningfully address it. Though the Tikhonov and TV-min methods can be well-parameterized to be robust to noise, they too have no mechanism to directly account for spatial blur. In standard workflows, blur removal typically requires a complex parameterization study, and is addressed as an independent pre-process. With LPD, such sources of measurement uncertainty can simply be learned from well-selected training data.

\begin{table}[H]
    \centering
        \caption{Comparing the mean PSNR, SSIM, and NMSE over 1250 tomographic reconstructions of the large $500\times 128$-pixel synthetic scenes using the Gridrec, Tikhonov, TV-Minimization, and the Learned Primal Dual methods.}
    \label{tab:parallel_results}
    \begin{tabular}{|l|c|c|c|}
       \hline 
        Method & PSNR & SSIM & NMSE  \\
         \hline                               
         Gridrec & 9.19 & 0.872 & 1.90E-1 \\
         Tikhonov & 11.2 & 0.901 & 1.02E-1 \\
         TV-Minimization & 19.8 & 0.942 & 3.31E-2\\
         Learned Primal Dual & 44.1 & 0.995 & 4.15E-3 \\
        \hline 
         
    \end{tabular}
\end{table}

\subsection{Cone-beam Modality}

Here, we construct and validate an LPD-based single-view tomographic reconstruction solver specific to the Scintillator EvAluation Laboratory (SEALab) at the Los Alamos Operations office of the NNSS. This test bench includes a Comet X-ray tube with an endpoint energy of 225keV operated at 3.5mA, a high-brightness GOS:Tb phosphor screen (DRZ-High), an Andor Neo Scientific Complementary Metal-Oxide-Semiconductor (sCMOS) camera, and an 85mm Zeiss lens operated at a fixed aperture of f/4. The distances between the source and detector, and source to the proposed axis of target symmetry were measured to be $\delta_{sd} = 203$cm and $\delta_{so} = 60.5$cm, respectively. The result is a radiographic region of interest (ROI) with a nominal geometric magnification $\delta_{sd}/\delta_{so} = 3.355$.

To characterize the noise and blur of the SEALab, we performed a small Monte Carlo study collecting 30 repeated radiographs of various diagnostic phantom targets. We utilized the step wedges to estimate noise levels as a function of photon transmission, and a Tungsten rollbar placed in a position of high geometric magnification (8.943) to estimate the X-ray source spot size. Transmission images of these scenes can be seen Figure \ref{fig:calibration}. 

\begin{figure}[H]
    \centering
    \includegraphics[width = 3.4in]{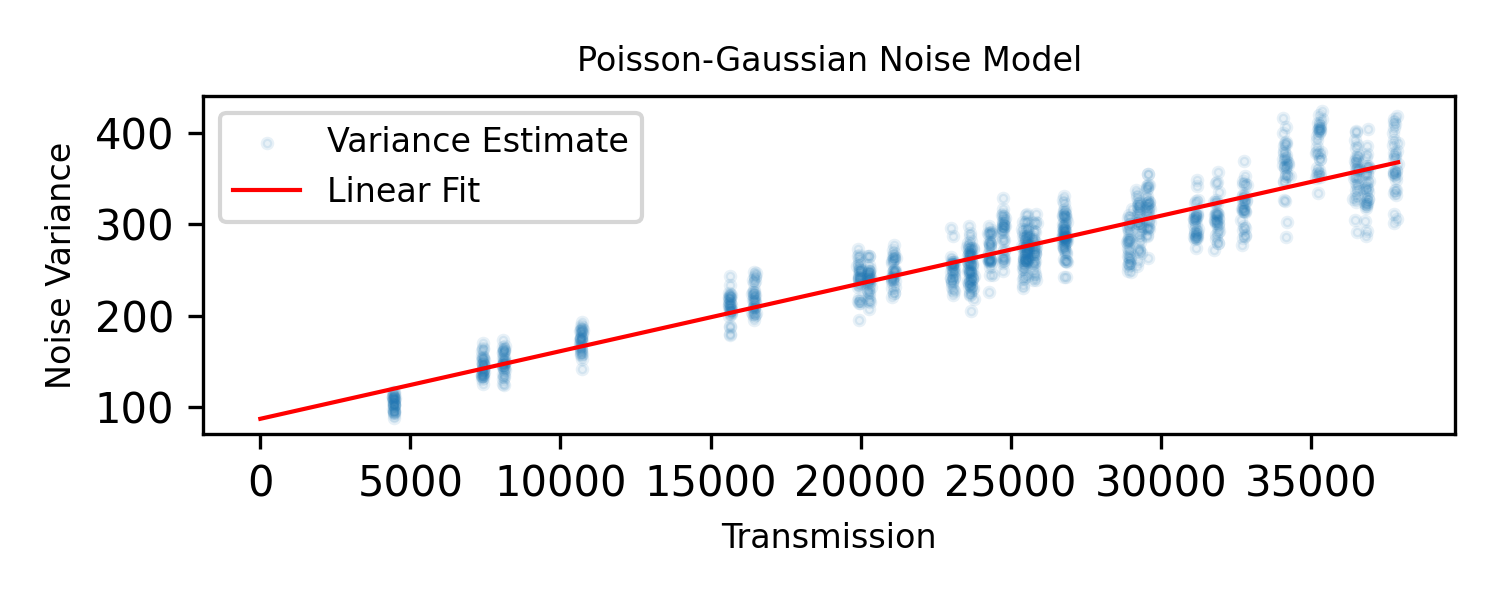}
    \caption{Local variance estimates were computed from 10x10-pixel blocks central to each of the calibration steps seen in Figure \ref{fig:calibration}. Each variance estimate is depicted by a semi-transparent blue marker. Radiography of the calibration target was captured and sampled 30 times. A linear fit was performed, as depicted by a red line.}
    \label{fig:PGN}
\end{figure}

An analysis of the resulting radiography determined that the noise for any observed pixel $d_{i,j}$ follows a combined Poisson-Gaussian process
$$\hat{d}_{i,j} = m p_{i,j} + g_{i,j}, \ \text{with} \ \ p_{i,j} \sim P(d_{i,j}), \ \ g_{i,j} \sim \mathcal{N}(\mu,\sigma^2).$$
We analyzed 30 sequential radiographs of the same diagnostic step wedge phantom seen in Figure \ref{fig:calibration}. For each step from each image, the mean and variance of a $10\times10$ pixel region was sampled. Following \cite{foi2008practical}, a linear fit was performed (see Figure \ref{fig:PGN}), establishing the forward noise model.

\begin{figure}[H]
    \centering
    \includegraphics[width = 3.4in]{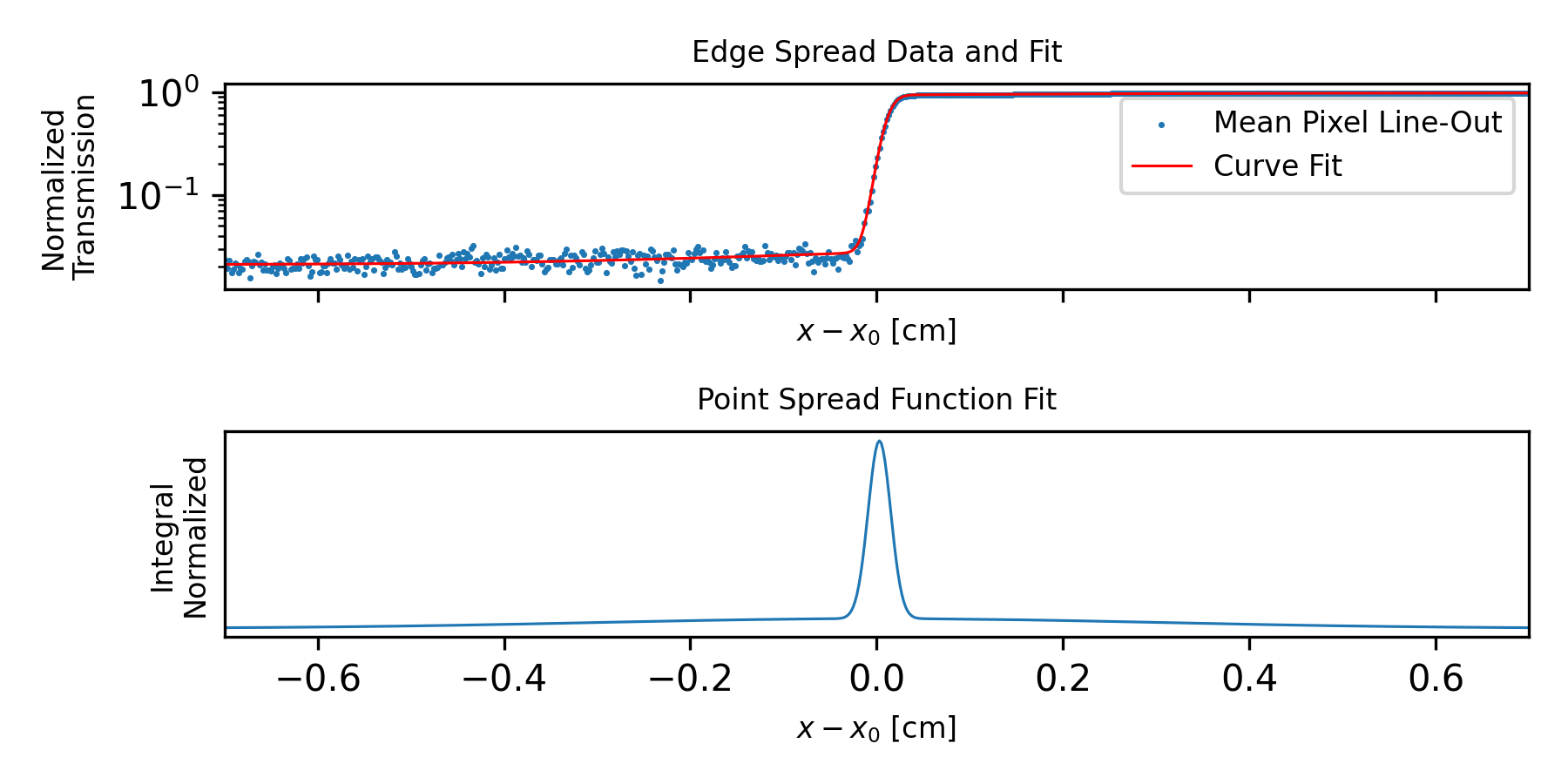}
    \caption{The top figure is a semi-log plot of a horizontal line-out of the $W$ rollbar, averaged over 30 radiographic exposures, and a regression fit to the ESF model in \eqref{eqn::esf}. The bottom shows a line-plot of the resulting normalized PSF that can be inferred from the regression weights fit to the ESF model.}
    \label{fig:esfpsf}
\end{figure}

Similarly, a model for the total blur of the system was estimated using a Tungsten rollbar, fitting a parameterized regressor to the edge-spread function (ESF) following \cite{schach2002using}. The resulting point-spread function (PSF) kernel $K$ is assumed to be radially symmetric, fully determined by the  parameters ${\bf a} = (a_1,\ldots,a_6)$ such that
$$K(r;{\bf a}) = a_1 e^{-(a_2 r)^2} + \frac{a_3 a_4^2}{(1 + (a_4 r)^2)^{3/2}} 
+ \frac{a_5 a_6^2}{(a_6^2 + r^2)^2},$$
and unit normalized. Given $K$, the associated model for the ESF follows from the Abel transform $\mathcal{A}$ \eqref{eqn::Abel}, which gives  
$$\frac{\text{d}}{\text{d}x} ESF(x; {\bf a}) = \mathcal{A}(K(r; {\bf a}))(x),$$ resulting in
\begin{multline} ESF(x;{\bf a}) = a_1\left(1 + \text{Erf}(a_2(x-x_0))\right) \\ + a_3\left(\frac{1}{2} + \frac{1}{\pi}\arctan\left(a_4(x-x_0)\right)\right) \\ + \frac{a_5}{2}\left( 1 + \frac{(x-x_0)}{\sqrt{a_6^2 + (x-x_0)^2}}\right),\label{eqn::esf}
\end{multline}
where $x_0$ denotes the estimated center of the sigmoid curve. The fit can be seen in Figure \ref{fig:esfpsf}.

Each synthetic target scene was constructed from random combinations of randomly-generated 3D axisymmetric objects. A collection of $10,000$ STL-formatted spheres, ellipsoids, cylinders, tori, annuli; either filled solid or assigned a uniformly-random shell thickness, were generated by Open CASCADE in GMSH \cite{opencascade,geuzaine2009gmsh}. In an effort to reduce data leakage, the first $8,000$ objects were reserved for the training set, while the remaining $2,000$ were withheld for the evaluation set. 

\begin{figure*}[t]
    \centering
    \includegraphics[width=5.5in]{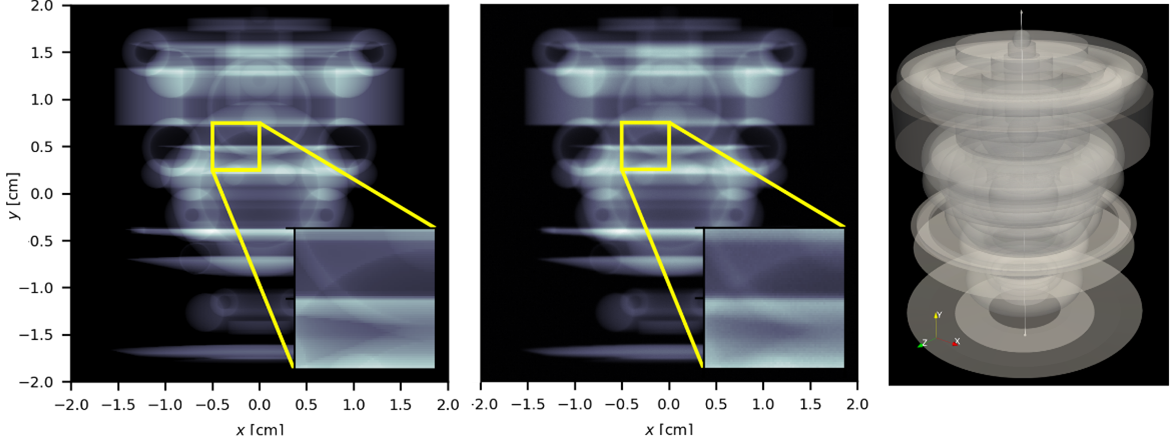}
    \caption{A perspective rendering of a surface mesh model used to generate randomized target scenes is shown on the right. The plots on the left and center respectively depict a ray-trace of the target scene imaged using gVXR with and without added blur and noise distortions.}
    \label{fig:SyntheticConeRadiography}
\end{figure*}

Each scene contains at least one, but up to forty 3D objects. Given that objects often overlap, we utilized volumetric Boolean operations in Open CASCADE to isolate each contiguous region. For simplicity, we assume that each region is made of Aluminum, but with a mass density selected uniformly-randomly between 0.1 and 20 $g/cm^3$. Lastly, for any particular random scene to be added to the dataset, no synthetic ray paths from the source to the detector can be fully obstructed. An example is depicted in Figure \ref{fig:SyntheticConeRadiography}.

Similar to the parallel beam case, our training data consists of pairs $(u, \mathcal{F}(u))$, which are initially generated with neither noise, nor the radiographic spot blur. The gVXR platform first performs a direct ray-trace using the Beer-Lambert law against models of the Bremsstrahlung source and DRZ scintillator. Computationally expensive effects such as scattering are ignored, but the modeled detector noise and source blur are added as post-processes.

\subsection{Training Details} A total of 6,250 pairs were generated, along the same 80/20 split, and $5\times$ augmentation, resulting in $25,000$ training pairs $(u_k,d_k)$. Each truth and distorted projection pair are $512 \times 512,$ and $512 \times 256$ respectively. In cone-beam acquisition, the strong inter-row coupling arising from source divergence prevents us from implementing similar model simplifications that were applied to the parallel-beam case. The higher degree of geometric coupling in addition to localized noise and blur make the recovery of fine structural details significantly more challenging. As a result, we trained our method against images the same size as the intended imaging scenario. Given the larger image spaces, our batch size was reduced to 128. We completed 650 epochs, with a final MSE loss of 5.45E-3.

\subsection{Results}
Mirroring the results found in the parallel-beam case, the LPD solver outperformed the included classical solvers; with similar caveats. We selected FDK \cite{feldkamp1984practical} as the unregularized baseline reconstruction example. This is a fast method that is similar to Gridrec, but can directly accommodate the cone-beam acquisition. As before, we include results from both the Tikhonov and TV-min methods, again identifying their respective regularization parameters $\lambda$ using a grid search of the evaluation data.

The LPD method again outperforms FDK, Tikhonov, and TV-min, but with a slightly tighter quantitative margin. This outcome can, in part, be attributed to a network optimization trade-off: time and resource constraints limit the overall number of training epochs, and the higher memory demands of the 512$\times$512 image space force a reduction in the training batch size. Smaller batches tend to introduce higher variance into the gradient estimates, ultimately resulting in slower convergence. In spite of this training limitation, the performance of LPD remains decisively higher than that of the classical methods. 

\begin{figure*}[t]
    \centering
    \includegraphics[width=6in]{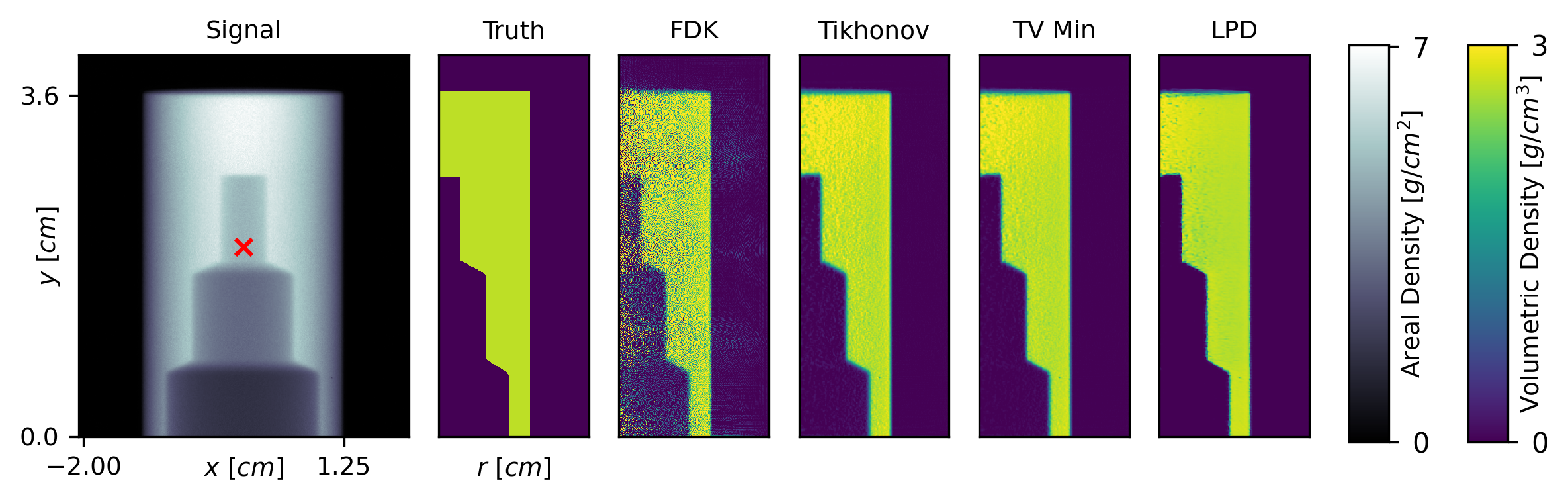}
\caption{From left to right, we start with a genuine radiograph of the cylindrically-symmetric target object, and a corresponding geometric estimate of the ground truth. The red $\times$ on the radiograph depicts the measured center of alignment. The four reconstructions that follow vary only the reconstruction methodology. The third image is computed using FDK, followed by Tikhonov and TV-min. The result from the LPD scheme is shown last. The horizontal and vertical axes are scaled to the central object plane, and translated to depict the scale of the target.}
    \label{fig:compare_with_direct_cone}
\end{figure*}

To demonstrate performance on real-world measurements, we present qualitative reconstructions of genuine flash radiography data collected at the SEALab. The target phantom consists of a machined aluminum cylinder, with an empty cavity drilled to various diameters. The original radiography (in absorption units) and a sliced cross-section of the idealized ground truth can be seen on the left side of Figure \ref{fig:compare_with_direct_cone}.

\begin{table}[H]
    \centering
        \caption{Comparing the mean PSNR, SSIM, and NMSE over 1000 tomographic reconstructions of the synthetic evaluation set, using the FDK, Tikhonov, TV-Minimization, and the Learned Primal Dual methods.}
    \label{tab:cone_results}
    \begin{tabular}{|l|c|c|c|}
       \hline 
        Method & PSNR & SSIM & NMSE  \\
         \hline                               
         FDK & 11.2 & 0.726 & 1.57E-1 \\
         Tikhonov & 23.2 & 0.910 & 6.53E-2 \\
         TV-Minimization & 29.1 & 0.913 & 4.31E-2\\
         Learned Primal Dual & 34.1 & 0.974 & 2.90E-2 \\
        \hline 
         
    \end{tabular}
\end{table}

Our characterizations of detector noise and spot blur were applied to the training data as a monolithic post-process. With genuine radiography, these quantities tend to vary on a variety of spatial scales. Additionally, neither imperfections in the scintillator, nor scattering were included. However, as seen in Figure \ref{fig:compare_with_direct_cone}, in spite of these limitations, LPD still performs qualitatively well compared the direct methods. It is our expectation that adding any combination of these features to the training data would further improve the results of a well-trained LPD method. 

\section{Discussion and Conclusions} \label{sec::conc}
The numerical demonstrations across both parallel-beam and cone-beam modalities  validated the LPD architecture as a robust solution for single-view tomographic reconstruction of axially-symmetric targets. Well-trained models surpassed traditional methods in challenging settings, demonstrating a clear capability to compensate for complex, unmodeled acquisition errors. 

We suspect that LPD is sufficiently adaptable to learn additional localized features (e.g. spatially-varying noise, scatter, and blur) without sacrificing adherence to the acquisition model constraints. Indeed, given that the forward and adjoint models ($\mathcal{A}$ and $\mathcal{A}^*$) exist outside of the shallow CNNs, we further suspect that small topological perturbations to the acquisition geometry may be accommodated. Future work should seek to benchmark the performance of LPD against subtle changes to radiographic magnification, and changes to the location and orientation of the axis of symmetry. Demonstrating this transfer learning pathway is the critical next step toward reducing the high computational cost associated with training high-resolution learned solvers for complex systems.


%

\section*{Acknowledgments}
This work was done by Mission Support and Test Services, LLC, under Contract No. DE-NA0003624 with the U.S. Department of Energy and the National Nuclear Security Administration’s Office of Defense Programs. DOE/NV/03624$\text{-}\text{-}$2280.

This paper is the direct result of several illuminating conversations that occurred at the \emph{Integrating Acquisition and AI in Tomography} workshop, hosted by the Lorentz Center at Leiden University, the Netherlands. The authors would also like to acknowledge the use of the High Performance Computing (HPC) resources and support provided by Livermore Computing (LC) at Lawrence Livermore National Laboratory.

\ifCLASSOPTIONcaptionsoff
  \newpage
\fi



\bibliographystyle{IEEEtran}
\bibliography{bibtex/bib/IEEEexample}
%

%








\end{document}